\documentclass [superscriptaddress,reprint,amsmath,amssymb,showkeys]{revtex4-2}
\usepackage{graphicx}% Include figure files
\usepackage{dcolumn}% Align table columns on the decimal point
\usepackage{bm}% bold math
\usepackage{color}
\usepackage{float}
\usepackage[utf8]{inputenc}
\usepackage[mathlines]{lineno}% Enable numbering of text and display math
\usepackage{comment}
\usepackage{amsmath}
\usepackage{xr}
\usepackage{xr-hyper}
\usepackage{hyperref}
\usepackage{soul,xcolor}
\soulregister\cite7
\soulregister\ref7
\setstcolor{blue}
\newcommand{\MPB}{MAPbBr$_3$}
\newcommand{\FPB}{FAPbBr$_3$}

\newcommand{\cmi}{cm$^{-1}$}
\newcommand{\degs}{$^\circ$}

\begin{document}

\title{Static and Dynamic Disorder in Formamidinium Lead Bromide Single Crystals}

% Affiliations; 
%"loosness=-1" - makes the affiliation appear in one line instead breaking it into two
\author{Guy Reuveni}
\affiliation{Department of Chemical and Biological Physics, Weizmann Institute of Science, Rehovot 76100, Israel;\looseness=-1}
\author{Yael Diskin-Posner}
\affiliation{Chemical Research Support, Weizmann Institute of Science, Rehovot 76100, Israel;}
\author{Christian Gehrmann}
\affiliation{Department of Physics, Technical University of Munich, 85748 Garching, Germany;}
\author{Shravan Godse}
\thanks{Present address: Mechanical Engineering Department, Carnegie Mellon University, Pittsburgh, Pennsylvania 15213, USA}
\affiliation{Department of Physics, Technical University of Munich, 85748 Garching, Germany;}
\author{Giannis G. Gkikas}
\affiliation{Department of Materials Science and Technology, University of Crete, Voutes Campus, Heraklion GR-70013, Greece;\looseness=-1}
\author{Isaac Buchine}
\affiliation{Department of Molecular Chemistry and Materials Science, Weizmann Institute of Science, Rehovot 76100, Israel;\looseness=-1}
\author{Sigalit Aharon}
\author{Roman Korobko}
\affiliation{Department of Chemical and Biological Physics, Weizmann Institute of Science, Rehovot 76100, Israel;\looseness=-1}
\author{Constantinos C. Stoumpos}
\affiliation{Department of Materials Science and Technology, University of Crete, Voutes Campus, Heraklion GR-70013, Greece;\looseness=-1}
\author{David A. Egger}
\affiliation{Department of Physics, Technical University of Munich, 85748 Garching, Germany;}
\author{Omer Yaffe}
\email{omer.yaffe@weizmann.ac.il}
\affiliation{Department of Chemical and Biological Physics, Weizmann Institute of Science, Rehovot 76100, Israel;\looseness=-1}

\begin{abstract}
We show that formamidinium lead bromide is unique among the halide perovskite crystals because its inorganic sub-lattice exhibits intrinsic local static disorder that co-exists with a well-defined average crystal structure.  
Our study combines THz-range Raman-scattering with single-crystal X-ray diffraction and first-principles calculations to probe the inorganic sub-lattice dynamics evolution with temperature in the range of 10 - 300~K.
The temperature evolution of the Raman spectra shows that low-temperature, local static disorder strongly affects the crystal's structural dynamics and phase transitions at higher temperatures.  

\end{abstract}

\maketitle

%\section*{Introduction}

Extensive research on lead-halide perovskites (APbX$_3$ where X=Cl, Br, I) has been primarily motivated by potential photovoltaic applications.\cite{gratzel2014light,zhou2014interface,lee2012efficient,wang2019stabilizing,kim2021comprehensive,dunlap2018synthetic,mitzi2019introduction,li2017chemically,tan2014bright,ricciardulli2021emerging,correa2017rapid,saliba2016cesium,correa2017promises} 
It has challenged the use of the primitive unit cell, together with the average crystal structure as determined by X-ray diffraction (XRD) measurements it represents for explaining certain properties of these crystals.
Specifically, we and others have shown that at sufficiently high temperatures, halide perovskites exhibit large amplitude (\textit{i.e.}, anharmonic) PbX$_6$ octahedral rotations and distortions.\cite{Ferreira2020,menahem2021,marronnier2018influence,bechtel2019finite,yaffe2017local,carignano2017critical,maughan2018anharmonicity}
Strongly anharmonic dynamic disorder implies that the actual crystal structure includes local motifs that are not captured when the material is represented by its average structure using the primitive unit cell.\cite{Zhu2022probing,Klarbring2020anharmonicity}
Notably, these local structural motifs were shown to impact the optoelectronic properties of lead-halide perovskites in a profound manner.\cite{zhao2020polymorphous,zhao2021effect,quarti2016structural,Gehrmann2022transversal}

\begin{figure}[h!]
\centering
\includegraphics[width=0.92\columnwidth]{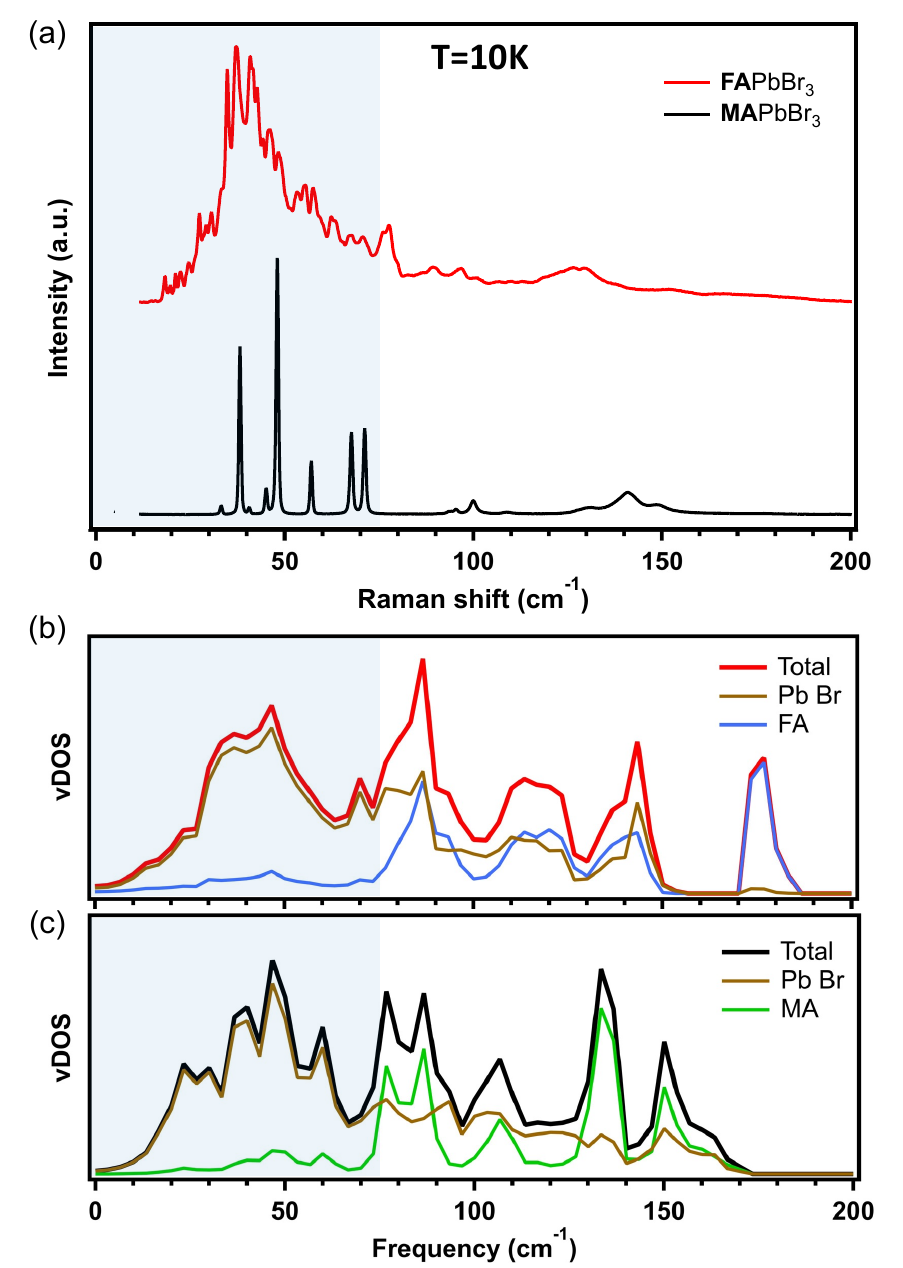}
\caption{(a) Raman-scattering spectra of \FPB\ (red) and \MPB\ (black) single crystals at 10~K.
The spectra are offset for clarity.
Vibrational density-of-states (vDOS) as well as decomposition of the vDOS for the organic and inorganic sublattices of (b) \FPB\ and (c) \MPB, as calculated by density functional theory.
Gray-shaded area marks the frequency range in which the vDOS predominantly features vibrations of the PbBr$_6$ framework rather than the organic cation.
}
\label{10K_Raman_and_DFT}
\end{figure}

\begin{figure*}
    \centering
    \includegraphics[width=\textwidth]{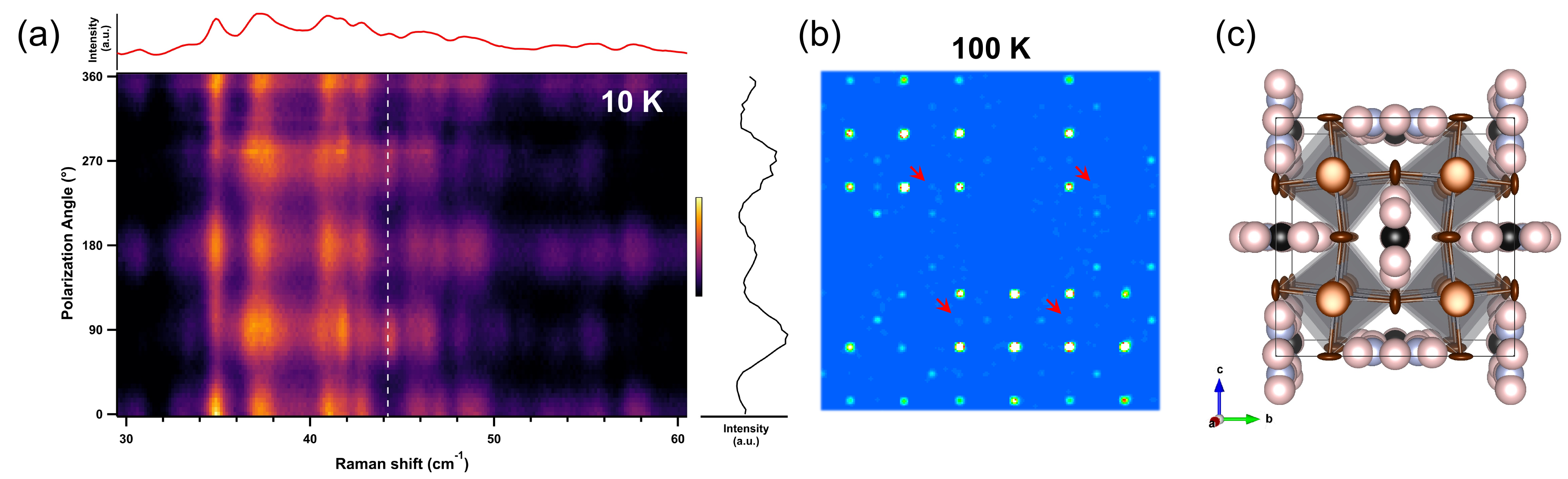}
    \caption{\FPB\ crystallography through polarization-orientation Raman-scattering and single-crystal X-ray diffraction.
    (a) False-color polarization-orientation Raman plot of \FPB\ at 10~K.
    The top panel shows part of the unpolarized Raman spectrum (as depicted in Fig.~\ref{10K_Raman_and_DFT}(a)), and the right panel shows the cross-section of the 44 \cmi\ peak (marked by a white dashed line), presenting its polarization-dependent intensity.
    Spectra were normalized to the highest peak; their intensities are represented by the color scale.
    (b) Precession image of \FPB\ at 100K (\textit{Immm} space group, (10-1) projection).
    Red arrows point at some non-indexed satellite reflections of very low intensity, corresponding to a larger (likely doubled) supercell.
    Bright reflections correspond to the indexed reflections of the \textit{Immm} space group (\#71).
    (c) Schematic representation of the 100~K crystal structure of \FPB\ obtained from single-crystal X-ray diffraction measurements.
    Gray, brown, blue, black, and pink spheres denote Pb, Br, N, C, and H atoms, respectively.
}
    \label{XRD_PO}
\end{figure*}

When the effect of dynamic disorder is negligible at sufficiently low temperatures, lead-halide perovskites based on methylammonium (MA) cation do not exhibit disorder.\cite{leguy2016,Poglitsch1987,wasylishen1985cation,singh2016effect}
Therefore, the representation of these crystals by a primitive unit cell can adequately describe the properties of these crystals. 
By contrast, recent studies of lead-halide perovskites with formamidinium (FA) as the A-site cation demonstrated high orientational disorder of the FA cation within the inorganic framework, even at cryogenic temperatures.\cite{Ferreira2020,maheshwari2019relation,mozur2019dynamical,taylor2018investigating,weber2018phase}
Since the FA cation is large, disordered, and interacts with the lead and the X-site ions electrostatically,\cite{Kabakova2018effect} it is likely to distort the perovskite inorganic framework.\cite{fabini2016dielectric,singh2020origin} 
At higher temperatures that are more relevant for device-operating conditions, the interplay of the orientational, static disorder of FA with the dynamic disorder in lead-halide perovskites may impact their optoelectronic properties but remains mostly unexplored to the best of our knowledge.

In this work, we investigate the possibility that static and dynamic disorder coexists in FA-based lead-halide perovskites by studying how the inorganic sublattice dynamics of \FPB\ differ from those of its MA-based counterpart \MPB.
We combine THz-range Raman-scattering with single-crystal X-ray diffraction (scXRD) and first-principles calculations to probe the structural dynamics of the PbBr$_6$ framework in \FPB\ as well as its temperature evolution (10-300~K) and compare it to the well-known case of \MPB.
We show that contrary to \MPB, the PbBr$_6$ framework in \FPB\ single crystals exhibits a significant degree of local static disorder despite having a well-defined average structure.
Our findings suggest that the static disorder at low temperatures is related to the bulky FA molecule and demonstrate that it augments the dynamic disorder present at higher temperatures in \FPB, which potentially has significant implications for the optoelectronic and thermal-stability properties of FA-based lead-halide perovskites.

%\section*{Results}

% \vspace{5pt}
\FPB\ and \MPB\ crystals were grown via the inverse temperature crystallization method and the antisolvent method, respectively.\cite{saidaminov2015retrograde,Ceratti2018}
Details regarding crystal growth, measurement apparatus, and parameters are given in the supporting information (SI).
Figure~\ref{10K_Raman_and_DFT}(a) presents a comparison between the unpolarized (\textit{i.e.}, summing over all polarization angles) THz-range Raman spectra of \FPB\ (red) and \MPB\ (black) single crystals at 10~K.
The features in the Raman spectrum of \MPB\ are well-resolved and agree with factor group analysis predictions based on the average crystal structure.\cite{yaffe2017local,cohen2022diverging,guo2017interplay}
In sharp contrast, the Raman spectrum of \FPB\ exhibits a strong background and contains more than 40 sharp peaks, higher than the expected number of Raman-active lattice modes ($\sim$18, depending on the specific space group and cell size used in factor group analysis\cite{kroumova2003bilbao}).
These striking differences between the two spectra indicate that \FPB\ is showing a high degree of static disorder of unknown origin.

One possible explanation for a large number of Raman peaks is that the relatively large FA cation\cite{becker2017,kieslich2014} may introduce additional modes in the THz frequency range. 
Molecular modes are not accounted for in factor group analysis, which only considers the space group of the average crystal (\textit{i.e.}, the organic cation is treated as a sphere).
To examine if the FA cation does indeed significantly change the low-frequency vibrational features in \FPB, we compute the vibrational density of states (vDOS) of cubic \FPB\ and \MPB\ using density functional theory (DFT) employing the VASP code (see SI for details).\cite{kresse1996efficient}
We calculated the vDOS of the cubic structures because, for \FPB, only the cubic phase was stable enough to perform phonon calculations. 
The total and decomposed (Pb-Br framework and organic cation) vDOS of \FPB\ and \MPB\ are presented in Fig.~\ref{10K_Raman_and_DFT}(b) and (c), respectively.
The comparison finds the overall appearance of the spectral features to be rather similar, with the exception that the vibrations associated with the rigid-body motion of the molecules, in the range between $\sim$125~cm$^{-1}$ and $\sim$180~cm$^{-1}$, exhibit some differences, in line with the different moments of inertia and molecular masses of FA and MA and previous findings reported for MAPbI$_3$ and FAPbI$_3$.\cite{Druzbicki2021cation}
Importantly, the results show that in \FPB\ the low-frequency part of the vDOS up to approximately 75~cm$^{-1}$ (shaded area), \textit{i.e.}, the region that contains the multitude of sharp Raman features in the experiments (\textit{cf.} Fig.~\ref{10K_Raman_and_DFT}(a)), predominantly stems from vibrations of the inorganic PbBr$_6$ framework.
Together with the similarity of the computed vDOS of \FPB\ and \MPB\ in this low-frequency region, it suggests that the large number of peaks observed in the Raman spectrum of \FPB\ at 10~K are rooted in significant distortions of the inorganic framework, which are absent in our phonon calculations of the cubic structure. 

To test if the multitude of peaks in the Raman spectra of \FPB\ at 10~K results from nanodomains that form during the cooling process, we performed polarization-orientation (PO) Raman measurements at 10~K (Fig.~\ref{XRD_PO}(a)).
We measure the change in Raman-scattering intensity as a function of the angle between the linear polarization of the excitation laser and an arbitrary axis in the plane of the crystal surface (see Fig.~\ref{SI_PO_setup}).
This is useful because the fluctuations in intensity as a function of the polarization angle reflect the average symmetry of the measured sample.\cite{menahem2021,sharma2020}
Despite exhibiting a broad spectrum that does not obey the expected Raman selection rules, \FPB\ evidently also features a periodic PO Raman dependence, similar to that of \MPB\ (Fig.~\ref{SI_PO_Raman_temperatures}).
These results indicate that our \FPB\ crystal was not fractured into nanodomains because the PO dependence of a multi-domain system would not exhibit any periodicity.
Therefore, the data presented in Fig.~\ref{XRD_PO}(a) highlight an inherent confluence of properties in \FPB, demonstrating local disorder as expressed in the unpolarized Raman spectrum while still being crystalline on average.

\begin{figure}[h!]
    \centering
    \includegraphics[width=\columnwidth]{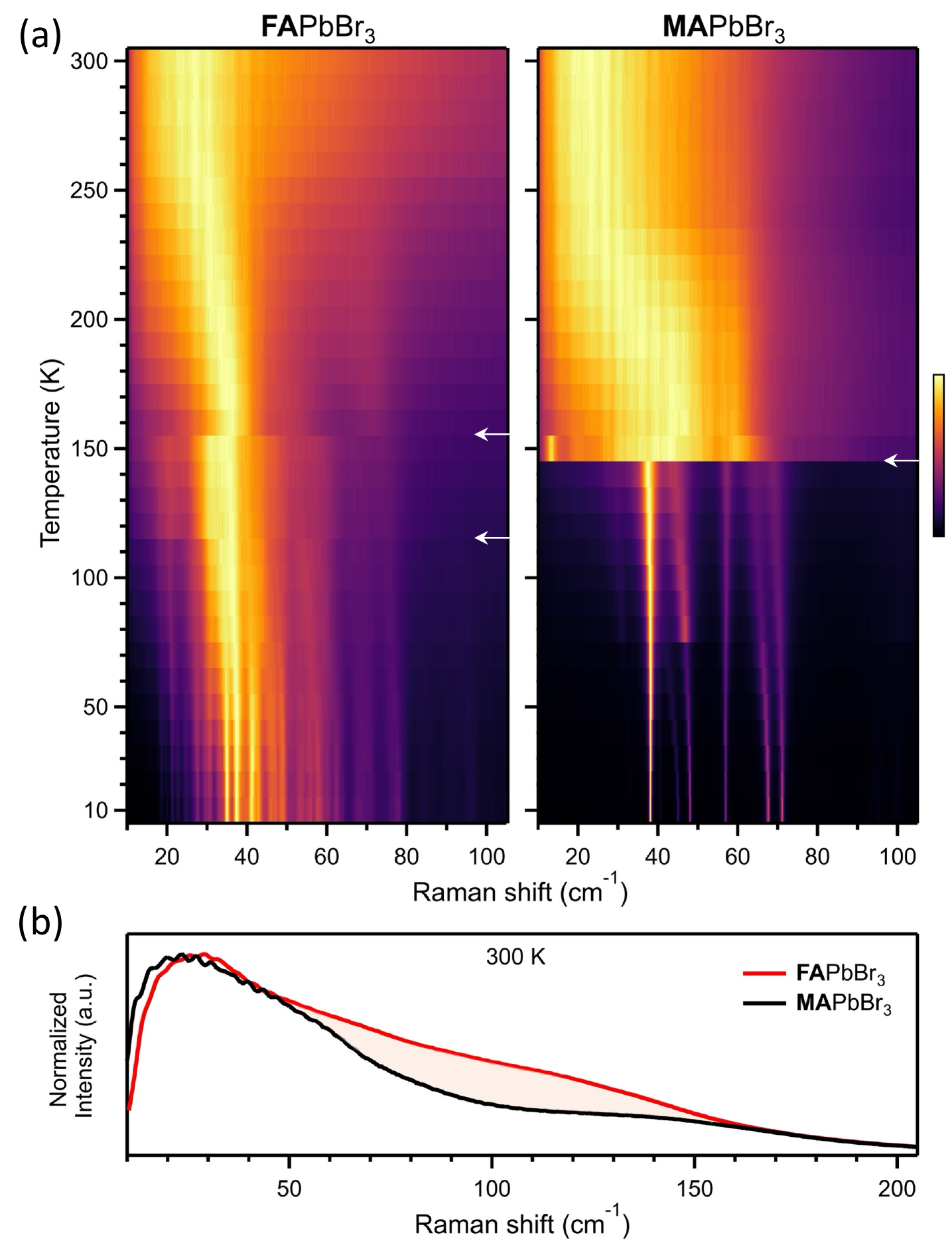}
    \caption{(a) Temperature-dependent Raman spectra from 10~K to 300~K, measured at 10~K intervals, of \FPB\ (left) and \MPB (right).
    All spectra were normalized to the highest peak, represented by brighter colors according to the color scale.
    White arrows point to observed phase transitions.
    (b) Raman spectra of the two crystals at 300~K, shaded area highlights the intensity difference.
    }
    \label{T_dep}
\end{figure}

To further investigate the crystallinity of our sample, we conducted scXRD measurements on the same \FPB\ crystal at 100~K. 
The comparison to the Raman data at 10~K is valid because, for this material, there are no reported phase transitions below 100~K.\cite{Govinda2018critical,schueller2018,keshavarz2019}
Figure~\ref{XRD_PO}(b) shows the precession image from scXRD at 100~K, where the reflections pattern appears as ordered, confirming that the inorganic PbBr$_6$ framework exhibits a well-defined average structure. 
Some low-intensity, non-indexed satellite reflections (marked with red arrows) appeared only in the 100~K data but not in any of the higher temperatures also measured in scXRD (see section~\ref{SI_sec_XRD}), suggesting the emergence of a supercell with doubled dimensions with respect to the \textit{Immm} unit cell.
The weak intensity of the satellite peaks precludes meaningful data integration regarding the supercell settings.
Specifically, while the refinement of the reflection pattern at 100~K showed a slight preference towards the orthorhombic \textit{Immm} space group over other space groups that were pointed out by group theoretical analysis,\cite{howard1998group,aleksandrov1976sequences,woodward1997octahedral} other independent structural refinements are ambiguous.
The ambiguity of the refinement, in addition to the Raman data shown in Fig.~\ref{XRD_PO}(a), indicates the presence of local disorder within the inorganic sublattice.
Altogether, the presence of these local disorder domains leads to an average structure of various local symmetries, each of which describes the XRD reflection pattern in an essentially equivalent manner.

Finally, we compare the temperature evolution of the structural dynamics of both lead-halide perovskite crystals.
Fig.~\ref{T_dep}(a) shows a false-color map representing the temperature-dependent Raman-scattering spectra of \FPB\ (left) and \MPB\ (right).
As temperature increases, the continuous broadening and red-shifting of the Raman features are apparent in both materials.
The data show that \MPB\ exhibits an order-disorder phase transition from orthorhombic to tetragonal around 145~K, similar to previous reports.\cite{onoda1990calorimetric,Poglitsch1987}
This phase transition is noticeable as a substantial broadening of the spectrum due to strongly anharmonic thermal fluctuations of the PbBr$_6$ octahedra,\cite{guo2017interplay,yaffe2017local} \textit{i.e.}, due to dynamic disorder.
\FPB\ shows a similar temperature evolution with apparent phase transitions at similar temperatures, with two noticeable spectral changes occurring around 115~K and 155~K (marked with white arrows), similar to the reported temperatures of structural phase transformations in \FPB.\cite{keshavarz2019}
However, these spectral changes are mild compared to what is observed in \MPB.
These differences are due to the fact that the Raman spectrum of \FPB\ is already very broad at low temperatures, as discussed above, to the extent that a phase transition is not resulting in further drastic changes.
Nonetheless, as temperature increases, dynamic disorder gradually becomes more significant also in this case, and accordingly, the Raman spectra of \FPB\ and \MPB\ become more alike.
In Fig.~\ref{T_dep}(b), we compare the Raman spectra of \FPB\ and \MPB\ at 300~K.
At this temperature, both spectra are dominated by dynamic disorder and exhibit diffused intensity at low frequency ($<$50~cm$^{-1}$).
The main difference between the spectra occurs between 50-170~cm$^{-1}$ (shaded area), where the relative intensity is significantly higher for \FPB. 
We hypothesize that this increased intensity indicates that the motions of the FA molecule in the inorganic cage distort the PbBr$_6$ framework while that of the MA molecule does not.
In light of the improved stability and self-healing properties of \FPB,\cite{juarez2019thermal,Ceratti2018,cahen2017self} it is worth investigating the role of the FA molecule for the dynamic disorder of the PbBr$_6$ framework in future experimental and theoretical work.

%\section*{conclusions}
To conclude, we used THz-range Raman-scattering, single crystal x-ray diffraction, and first-principles calculations to show that contrary to \MPB, the PbBr$_6$ framework in \FPB\ is intrinsically disordered while having a well-defined average structure at cryogenic temperatures.
Consequently, the local structure does not coincide with the average structure, and a supercell accounting for the FA cation orientational disorder, together with the resulting distortions of the inorganic framework, may better describe the physical situation of the material.
When temperature increases, the dynamic disorder becomes more significant, leading to a high resemblance between \MPB\ and \FPB, distinct only by the higher static disorder of the latter.
\FPB\ thus serves as an intriguing system that combines long-range crystal order with inherent local, static disorder, potentially being a key to describing the complex behavior of this unique perovskite.

\vspace{8mm}
\section*{acknowledgments}
The authors thank David Cahen for fruitful discussions and Lior Segev for software development.
G.R. acknowledges the support for this research by the Weizmann Institute Sustainability and Energy Research Initiative.
O.Y. acknowledges funding from European Research Council (850041 — ANHARMONIC).
D.A.E. acknowledges funding from the Alexander von Humboldt Foundation within the framework of the  Sofja Kovalevskaja Award, the Technical University of Munich - Institute for Advanced Study (grant Agreement No. 291763), and the Deutsche Forschungsgemeinschaft (EXC 2089/1 - 390776260).

%\input{main.bbl}
% \bibliography{literature.bib}

%%%%%%%%%%%%%%%%%%%%%%%%%%%%%%%%%%%%%%%%%%%%%%%%%%%%%%%%%%%%%%%%%%%%%%%%%%%%%%%%%%%%%%%%%%%%%%%%%%%%%%%%%%%%%%%%%%%%%%%%%%%%%%%%%%%%%%%%%%%%%%%%%%%%%%%%%%%%%%%%%%%%%%%%%%%%%%%%%%%%%%%%%%%%%%%%%%%%%%%%%%%

\clearpage
\onecolumngrid

\centering{{\Huge Supporting Information}} \\ 
\centering{\textbf{Static and Dynamic Disorder in Formamidinium Lead Bromide Single Crystals}}

\renewcommand{\thepage}{S\arabic{page}}  
\renewcommand{\thesection}{S\arabic{section}}   
\renewcommand{\thesubsection}{S\arabic{section}.\alph{subsection}} 
\renewcommand{\thetable}{S\arabic{table}}   
\renewcommand{\thefigure}{S\arabic{figure}}
\renewcommand{\theequation}{S\arabic{equation}}
\setcounter{page}{1}
\setcounter{figure}{0}
\setcounter{equation}{0}

\section{Crystal synthesis procedures}

\subsection{\FPB}
        %Crystallized by Isaac Buchine
        \FPB\ was crystalized in the inverse temperature crystallization method to induce high-quality crystallization.\cite{saidaminov2015retrograde} FABr (1.1M) and PbBr$_2$ (1.0M) were mixed in a 1:1 ratio of $\gamma$-Butyrolactone (GBL) and Dimethylformamide (DMF). The precursor solution is stirred for 2 hours at room temperature and is then filtered directly into a crystallization plate. The plate is tightly covered with aluminum foil and placed into a preheated oven at 50\degs C. The oven’s temperature is then ramped slowly up to 80\degs C over 5 hours. Upon crystallization, the plate is removed, and crystals are dried quickly to prevent them from dissolving as a result of ambient temperature.
\subsection{\MPB}
         %Crystallized by Davide Ceratti, according to the process from his paper.
     \begin{figure*}[h!]
        \centering
        \includegraphics[scale=0.5]{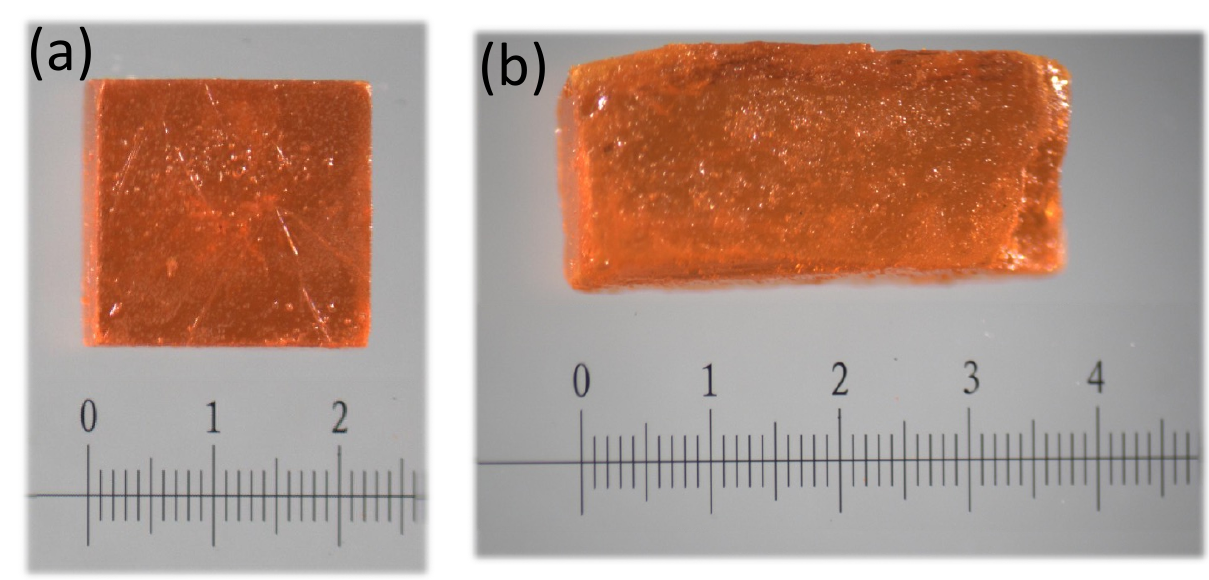}
        \caption{Microscope images of (a) \FPB\ and (b) \MPB\ single crystals.
                    The images were taken a long time after the measurements; thus, they are not transparent as fresh crystals.
                    Scale marks are in millimeters.}
        \label{SI_crystals_image}
    \end{figure*}

        \MPB\ crystals were synthesized by the antisolvent method.\cite{Ceratti2018}
        MABr (Dyesol) and PbBr$_2$ (Dyesol) were dissolved in DMF (N,N-dimethylformamide (Aldrich)) at the same time to obtain a 0.88 M solution in MA$^+$ and a 0.80 M solution in Pb$^{2+}$ (mixing at room temperature and in ambient air (RH 45\%)).
        It is essential to dissolve both chemicals in the same liquid because the solubility of PbBr$_2$ is increased by the presence of MABr.
        The solution was then filtered and put in an open vial. The open vial was placed in a wider bottle containing ethyl acetate (Bio-Lab) in excess.
        % fig.~\ref{SI_crystals_image}).
        The DMF solution, initially occupying half the available volume of the vial, absorbs ethyl acetate, which is an antisolvent for the halide perovskite.
        Crystals started to form and grow until the DMF- ethyl acetate-enriched solution level occupied all the available volume in the vial (between 24 h and 48 h in our case).
        The crystals were then extracted from the solution, carefully dried with blotting paper, washed with ethyl acetate, and dried again.

\section{\label{SI_sec_XRD}X-ray diffraction}
    The crystal structure and symmetry of the \FPB\ were determined by single-crystal X-ray diffraction (scXRD).
    The data and the refined structures can be found in the CIFs attached to the SI.
    % Overall, there was no significant preference for one symmetry refinement over the others,
    Albeit slightly higher refinement parameters resulted from an orthorhombic \textit{Immm} (\#74) space group, all space groups that are presented here have similarly described the reflection patterns.
    This ambiguity suggests a high lattice disorder that leads to an average structure combining various crystal symmetries.  
    
\subsection{Measurement procedure}
\begin{figure*}
        \centering
        \includegraphics[scale=0.6]{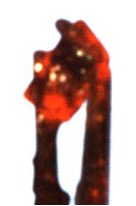}
        \caption{The mounted \FPB\ crystal.}
        \label{SI_xrd_crystal_image}
    \end{figure*}

    The synthesized \FPB\ crystal was broken into small pieces, then immersed in a small amount of nail polish and mounted onto a broken Mitogen loop at room temperature (see fig.~\ref{SI_xrd_crystal_image}).
    Measurements were performed on a \FPB\ single crystal starting at 300~K; then, the crystal was cooled down to 100~K at a rate of $1~K/min$.

    The cooling descent paused at 250~K, 200~K, and 150~K for data collection. A 5-minute wait before starting each measurement allowed the crystal to adjust to the temperature completely. 
    All data were collected on Rigaku Synergy-S dual source diffractometer equipped with Dectris Pilatus3 R CdTe 300~K detector and microfocus with AgK$\alpha$ ($\lambda$=0.56087~Å), with $\omega$ scans.
    This specific wavelength and detector setting are the most equipped in measuring these materials today.
    Data were collected to a very high resolution and redundancy.
    We have performed two independent structural solution and refinement processes.
    These two independent refinement solutions are presented herein.
    
\subsection{Traditional refinement}
    Data were integrated and processed according to the highest and most suitable Laue symmetry with CrysAlis$^{PRO}$, and a Gaussian absorption correction was applied. 
    Structures were solved using SHELXT\cite{Sheldrick2015shelxt} and further refined with SHELXL\cite{Sheldrick2015shelxl} by full-matrix least-squares refinement based on F$^2$. The major contributors to the diffraction scattering are the Pb and Br atoms, whose position is quickly determined.
    The lead bromide sublattice, composed of excellent contributors to the diffraction scattering pattern, defines the crystal symmetry.
    The formamidinium molecule in the cavity is probably disordered and has a negligible contribution to the diffraction scattering pattern.
    Its atoms' contribution to the diffraction scattering pattern is extremely small, even at low temperatures such as 100~K.
    A small undefined electron density blob exists in the electron density map, which is contributed by one disordered formamidinium molecule sitting on a symmetry fold, even with data sets measured to a very high resolution, such as 0.4~Å.
    Therefore, the symmetry and space group are defined exclusively by the diffraction scattering from the lead and bromine atoms.
    
    Since the bromine atom's contribution to the diffraction pattern is substantial, the shift from the special position is genuine and followed by a dramatic decrease in the R factor during the refinement.
    The shift in the bromine atom position between the 150~K to the 100~K measurements changes the lattice conformation, which might indicate a phase transition.
    The refinement was completed by refining a formamidinium fragment into the remaining electron density.
    The fragment belonged to the TALTAE structure from the CCDC.

\subsection{Group-subgroup based refinement}
\begin{figure*}[hb]
    \centering
    \includegraphics[scale=0.5]{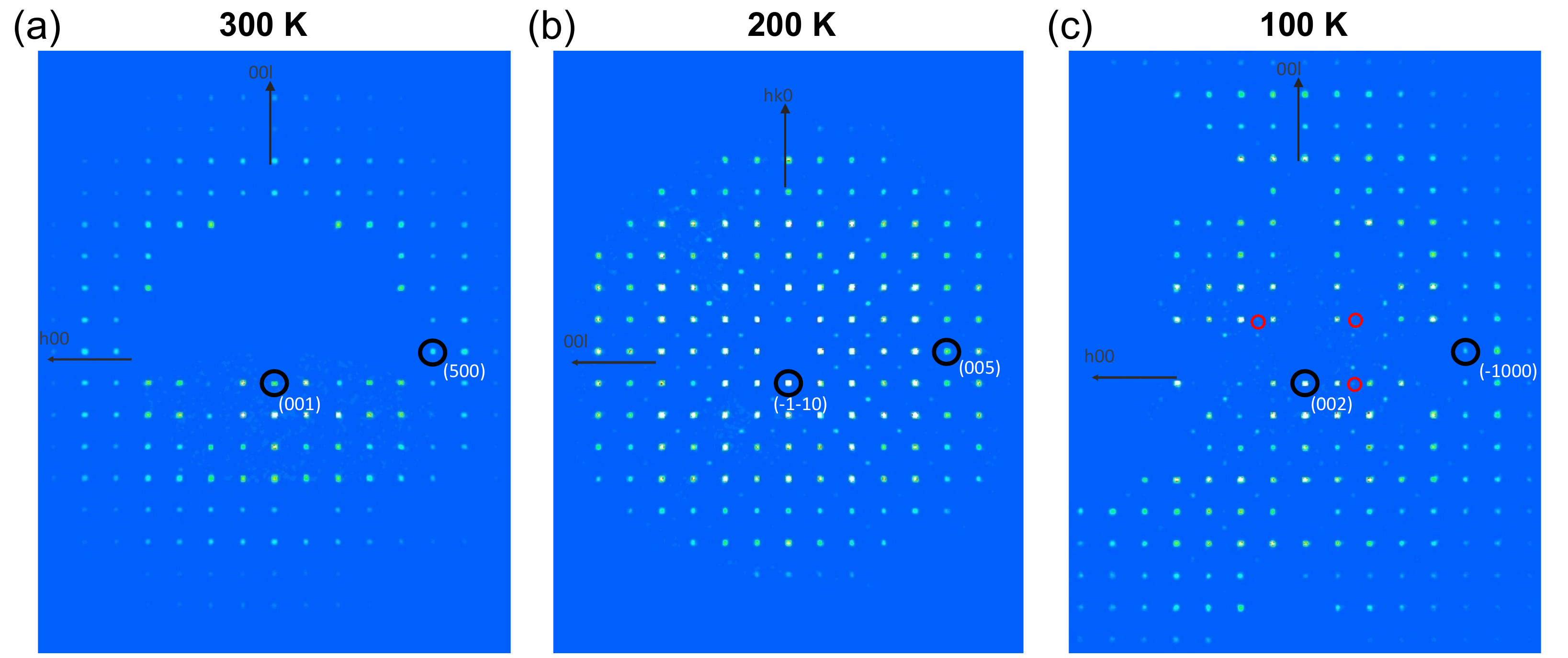}
    \caption{Precession images of \FPB\ at (a) 300K (\textit{Pm-3m} space group, (10-1) projection), (b) 200K (\textit{P4/mbm} space group, (221) projection), and (c) 100K (\textit{Immm} space group, (10-1) projection).
    Red circles represent non-indexed satellite reflections of very low intensity, suggesting the emergence of a large supercell.}
    \label{SI_XRD_precession}
\end{figure*}

    The second independent structural refinement set focuses on the three main space groups indicated by group theoretical analysis,\cite{howard1998group,aleksandrov1976sequences,woodward1997octahedral} at 100~K.
    Crystallographic data specific for the 100~K data set are summarized in tables~\ref{tbl:XRD_FPB100K}-~\ref{tbl:XRD_bond_angles}.\cite{Petricek2014crystallographic,Petricek2016crystallographic}
    Raw data were indexed with a variety of possible supercells of the primitive \textit{P4/mbm} space group that produced reasonably good indexing of the observed reflections ($>$90\%).
    Each indexing attempt was accompanied by a separate integration process (CrysAlis$^{PRO}$ platform), resulting in different sets of hkl files that were subsequently used to solve the structure (Superflip, integrated with Jana2006) and subsequently refine it (Jana2006).
    We observe a good agreement with all three possible orthorhombic space groups deriving from the tetragonal $\beta$-phase (\textit{P4/mbm} space group), namely \textit{Pbnm}, \textit{Cmcm}, and \textit{Immm}, and find a preferable agreement with the latter.
    Figure~\ref{SI_XRD_precession} presents precession images for 300~K, 200~K, and 100~K, demonstrating the ordered reflection pattern and approving our sample's crystallinity.

    Fig.~\ref{XRD_PO}c in the main article portrays the orthorhombic \textit{Immm} structure, which demonstrates through thermal ellipsoids how the bromine atoms (brown spheres) are highly shifted from their special Wyckoff position, compared to other atoms in the structure.
    Some non-indexed satellite reflections (red circles in fig.~\ref{SI_XRD_precession}(c)) appeared only in the 100~K data, which may suggest the emergence of a large supercell.
    Refining the structure as a supercell is a different yet reasonable approach that further implies the existence of the inherent disorder.
    That is because a supercell can include many different cation orientations alongside the inorganic framework distortions.
    This suggests that the inorganic framework of \FPB\ deforms in a very different manner compared to that of \MPB, possibly driven by the planar nature of the FA cation, which tends to orient in discrete directions along the three orthogonal axes, in contrast to MA which tends to orient along the body diagonals of the cage.\cite{Poglitsch1987}
    
\newpage 

 Temperature-dependent crystal data and structure refinement for \FPB\ - 1$^{st}$ independent refinement
 \begin{table}[t]
     \caption{Temperature dependent crystal data and structure refinement for \FPB\ - 1$^{st}$ independent refinement}
     \label{tbl:XRD_FPB_Tdep}
     \center
    %  \vspace{-0.4cm}
 %     \setlength{\arrayrulewidth}{.3em}
 % \bigskip
   \begin{tabular*}{\textwidth}{@{\extracolsep{\fill}}llllll}
     \hline 
%   \hlinewd{2pt}
    %  \vspace{0.1cm}
     Temperature & 300~K  & 250~K  & 200~K  & 150~K  & 100~K \\
     \hline 
    %  \vspace{0.2cm}
    Chemical Formula & HC(NH$_2$)$_2$PbBr$_3$ & HC(NH$_2$)$_2$PbBr$_3$ & HC(NH$_2$)$_2$PbBr$_3$ & HC(NH$_2$)$_2$PbBr$_3$ & HC(NH$_2$)$_2$PbBr$_3$\\ % \addlinespace[4pt]
     Formula weight & 492 & 492 & 492 & 492 & 492\\ % \addlinespace[4pt]
     Crystal system & cubic & cubic & cubic & cubic & cubic\\ % \addlinespace[4pt]
     Space group & \textit{Pm-3m} & \textit{Pm-3m} & \textit{Pm-3m} & \textit{Pm-3m} & \textit{Pm-3m}\\ % \addlinespace[4pt]
     a (Å) & 5.98010(10) & 5.97200(10) & 5.95470(10) & 5.93280(10) & 5.91992(4) \\ % \addlinespace[4pt]
     Volume~(\mbox{\normalfont\AA}$^{3}$) & 213.858(11) & 212.990(11) & 211.145(11) & 209.140(11) & 207.466(4)\\ % \addlinespace[4pt]
     Z & 1 & 1 & 1 & 1 & 1\\ % \addlinespace[4pt]
     Calculated Density ($g/{cm^3}$) & 3.820 & 3.836 & 3.869 & 3.906 & 3.938\\ % \addlinespace[4pt]
     $\mu\ (mm^{-1})$ & 18.134 & 18.208 & 18.367 & 18.543 & 18.693\\
     & \\
     % \multirow{}{}{Index ranges}
     & -12$\leq$ h$\leq$12 & -7$\leq$ h$\leq$7 & -7$\leq$ h$\leq$7 & -7$\leq$ h$\leq$7 & -12$\leq$ h$\leq$12\\
     Index ranges & -12$\leq$ k$\leq$12 & -7$\leq$ k$\leq$7 & -7$\leq$ k$\leq$7 & -7$\leq$ k$\leq$7 & -12$\leq$ k$\leq$12\\
     & -12$\leq$ l$\leq$12 & -7$\leq$ l$\leq$7 & -7$\leq$ l$\leq$7 & -7$\leq$ l$\leq$7 & -12$\leq$ l$\leq$12\\ % \addlinespace[4pt]
     & \\
     No. of reflection (unique) & 24055(243) & 5236 (78) & 5198 (78) & 5176 (76) & 23410 (233)\\ % \addlinespace[4pt]
     $\theta$$_{max}$(\degs) & 35.053 & 21.700 & 21.766 & 21.654 & 34.917\\ % \addlinespace[4pt]
     R$_{int}$ & 0.0418 & 0.0761 & 0.0712 & 0.833 & 0.0659\\ % \addlinespace[4pt]
     Completeness to  $\theta\ (\%)$	& 1.000	& 1.000	& 1.000	& 1.000	& 1.000\\ % \addlinespace[4pt]
     Data / restraints / parameters & 243 / 12 / 25 & 78 / 3 / 13 & 78 / 3 / 13 & 76 / 3 / 13 & 233 / 12 / 26\\ % \addlinespace[4pt]
     Goodness-of-fit on $F^2$ & 1.093 & 1.277 & 1.372 & 1.298 & 1.296\\ % \addlinespace[4pt]
     Final $R_1$ and w$R_2$ indices [I$>2\sigma$(I)] & 0.0145, 0.0336 & 0.0184, 0.0581 & 0.0291, 0.0848 & 0.0262, 0.0815 & 0.0086, 0.0176\\ % \addlinespace[4pt]
     $R_1$ and w$R_2$ indices [all data] & 0.0159, 0.0339 & 0.0184, 0.0581 & 0.0291, 0.0848 & 0.0262, 0.0815 & 0.0086, 0.0176\\
     & \\
     & \\
     & \\ 

   \end{tabular*}
 \end{table}

 % Structure parameters - 1$^{st}$ independent refinement
 \begin{table}[h]
     \caption{Structure parameters - 1$^{st}$ independent refinement}
     \label{tbl:XRD_FPB_parameters}
     \center
     \vspace{-0.3cm}
   \begin{tabular*}{\textwidth}{@{\extracolsep{\fill}}llllll}
     \hline 
     Temperature & 300~K  & 250~K  & 200~K  & 150~K  & 100~K \\ \hline
     Pb(1) - Br(1) Bond length (Å) & 2.99005(5) & 2.98600(5) & 2.97735(5) & 2.96790(5) & 2.97671(16)\\
     & \\
     % \multirow{}{}{Pb(1) (x,y,z)} 
      & 0.500000 & 0.500000 & 0.500000 & 0.500000 & 0.500000\\
      Pb(1) (x,y,z) & 0.500000 & 0.500000 & 0.500000 & 0.500000 & 0.500000\\
      & 0.500000 & 0.500000 & 0.500000 & 0.500000 & 0.500000\\
      & \\
     % \multirow{}{}{Br(1) (x,y,z)} 
      & 0.000000 & 0.000000 & 0.000000 & 0.000000 & 0.4467(2)\\
      Br(1) (x,y,z) & 0.500000 & 0.500000 & 0.500000 & 0.500000 & 0.500000\\
      & 0.500000 & 0.500000 & 0.500000 & 0.500000 & 0.000000\\
     \hline
    
   \end{tabular*}
 \end{table}

 % Crystal data and structure refinement for \FPB\ at 100~K - 2$^{nd}$ independent refinement
 \begin{table}[t]
     \caption{Crystal data and structure refinement for \FPB\ at 100~K - 2$^{nd}$ independent refinement}
     \label{tbl:XRD_FPB100K}
     \center
    %  \vspace{-0.4cm}
 %     \setlength{\arrayrulewidth}{.3em}
 % \bigskip
 %   \begin{tabularx}{\textwidth}{@{\extracolsep{\fill}}Xlll}
 %   \begin{tabularx}{\linewidth}{Xlll}
 %   \begin{tabularx}{\textwidth}{p{0.19\linewidth} p{0.2\linewidth}| p{0.2\linewidth} |p{0.2\linewidth} }
 % \resizebox{\textwidth}{!}
%  \footnotesize
   \begin{tabular*}{\textwidth}{@{\extracolsep{\fill}}lc|c|c}
   \hline
     Chemical Formula & \multicolumn{3}{c}{HC(NH$_2$)$_2$PbBr$_3$}\\% \addlinespace[2pt]
     Formula weight & \multicolumn{3}{c}{492}\\% \addlinespace[2pt]
     Temperature & \multicolumn{3}{c}{100~K}\\% \addlinespace[2pt]
     Wavelength & \multicolumn{3}{c}{0.56087~Å}\\% \addlinespace[2pt]
     Crystal system & \multicolumn{3}{c}{Orthorhombic}\\% \addlinespace[2pt]
     Space group & $\textit{Immm}$ & $\textit{Pbnm}$ & $\textit{Cmcm}$\\ %\hline
     & & & \\
     % \multirow{}{}{Unit cell dimensions} 
     & a=11.8423(3)~Å, & a=8.3673(2)~Å, & a=11.8432(3)~Å,\\
     & $\alpha$=90\degs & alpha=90\degs & alpha=90\degs \\
     Unit cell dimensions & b=11.8411(3)~Å, & b=8.3759(2)~Å, & b=11.8347(3)~Å,\\
     & $\beta$=90\degs & $\beta$=90\degs & $\beta$=90\degs \\
     & c=11.8346(3) ~Å, & c=11.8445(3)~Å, & c=11.8406(3)~Å,\\
     & $\gamma$=90\degs & $\gamma$=90\degs & $\gamma$=90\degs \\ %\hline
     & & & \\
     Volume~(\mbox{\normalfont\AA}$^{3}$) & 1659.52(7) & 830.11(4) & 1659.59(7) \\ % \addlinespace[2pt]
     Z & 8 & 4 & 8\\ % \addlinespace[2pt]
     Calculated Density ($g/{cm^3}$) & 3.9382 & 3.9365 & 3.938 \\ % \addlinespace[2pt]
     Absorption coefficient ($mm^{-1}$) & 18.793 & 18.785 & 18.792\\ % \addlinespace[2pt]
     F(000) & 1696 & 848 & 1696\\ % \addlinespace[2pt]
     Crystal size (mm) & 0.143x0.089x0.023 & 0.143x0.089x0.023 & 0.143x0.089x0.023\\ % \addlinespace[2pt]
     $\theta$ range for data collection & 2.71 to 35\degs & 2.35 to 35.21\degs & 2.35 to 35.23\degs\\ 
     % \multirow{}{}{Index ranges}
     & & & \\
     & -22$\leq$ h$\leq$24, & -16$\leq$ h$\leq$16, & -22$\leq$ h$\leq$24, \\
     Index ranges & -23$\leq$ k$\leq$21, & -17$\leq$ k$\leq$14, & -24$\leq$ k$\leq$22,\\
     & -24$\leq$ l$\leq$21 & -24$\leq$ l$\leq$22 & -23$\leq$ l$\leq$23\\
     & & & \\
     Reflections collected & 23768 & 22648 & 23681\\ % \addlinespace[2pt]
     Independent reflections	& 3960 [R$_{int}$ = 0.0502]	& 3612 [R$_{int}$ = 0.0511]	& 3959 [R$_{int}$ = 0.0506]\\ % \addlinespace[2pt]
     Completeness & 98\% (to $\theta$ = 34.85\degs) & 98\% (to $\theta$ = 23.89\degs) & 98\% (to $\theta$ = 34.85\degs) \\ % \addlinespace[2pt]
     & & & \\
     Refinement method & \multicolumn{3}{c}{Full-matrix least-squares on F$^2$}\\ % \addlinespace[2pt]
     & & & \\
     Data / restraints / parameters & 3960 / 4 / 39 & 3612 / 2 / 28 & 3959 / 2 / 36 \\ % \addlinespace[2pt]
     Goodness-of-fit & 1.63 & 2.50 & 2.68\\ % \addlinespace[2pt]
     Final R indices [I$>2\sigma$(I)] & R$_{obs}$=0.0425, wR$_{obs}$=0.0914 & R$_{obs}$=0.0642, wR$_{obs}$=0.1492 & R$_{obs}$=0.0660, wR$_{obs}$=0.1578\\
     & & & \\
     R indices [all data] & R$_{all}$ = 0.0763, & R$_{all}$ = 0.1052,& R$_{all}$ = 0.1045,\\
     & & & \\
     & wR$_{all}$ = 0.0990 & wR$_{all}$ = 0.1588 & wR$_{all}$ = 0.1665\\
     %\hline
     & & & \\
     % \multirow{}{}{Domain fractions}
      & \#1: 27.6(4)\% & \#1: 100\% & \#1: 55.9(5)\%\\
      & (1 0 0 0 1 0 0 0 1) & (1 0 0 0 1 0 0 0 1) & (1 0 0 0 1 0 0 0 1)\\
      Domain fractions & \#2: 38.8(2)\% & & \#2: 44.1(5)\% \\
      & (0 0 1 0 -1 0 1 0 0) &  & (0 -1 0 -1 0 0 0 0 -1)\\
      & \#3: 0.335(3) & & \\
      & (0 1 0 1 0 0 0 0 -1) & & \\
      %\hline
     & & & \\
     Largest diff. peak and hole & 10.27 and -6.78 e·Å$^{-3}$ & 7.55 and -6.68  e·Å$^{-3}$ & 7.69 and -11.39 e·Å$^{-3}$ \\
     \hline
     \multicolumn{4}{c}{R = $\frac{\Sigma||F_o|-|F_c||}{\Sigma|F_o|}$, wR = $\Big(\frac{\Sigma[w(|F_o|^2 - |F_c|^2)^2]}{\Sigma[w(|F_o|^4)]}\Big)^{1/2}$  and w=$\frac{1}{(\sigma^2(I)+0.0004I^2)}$}
 
   \end{tabular*}%}
 \end{table} 
 \newpage

 % Atomic coordinates ($x10^4$) and equivalent isotropic displacement parameters (Å$^2$x$10^3$) for \FPB\ at 100~K with estimated standard deviations in parentheses.
 \begin{table}[h]
     \caption{Atomic coordinates ($x10^4$) and equivalent isotropic displacement parameters (Å$^2$x$10^3$) for \FPB\ at 100~K (\textit{Immm} space group) with estimated standard deviations in parentheses.}
     \label{tbl:XRD_coordinates}
     \center
     \vspace{-0.3cm}
   \begin{tabular*}{0.7\textwidth}{@{\extracolsep{\fill}}lccccc}
     \hline 
     Label & x & y & z & Occupancy & U$_{eq}$*\\
     \hline
     Pb(1) & 5000 & 7501(1) & 7499(1) & 1 & 12(1)\\
     Br(1) & 5000 & 5000 & 7773(2) & 1 & 41(1)\\
     Br(2) & 7500 & 7500 & 7500 & 1 & 78(2)\\
     Br(3) & 5000 & 7247(3) & 5000 & 1 & 53(1)\\
     Br(4) & 5000 & 7746(3) & 10000 & 1 & 61(2)\\
     Br(5) & 5000 & 10000 & 7297(3) & 1 & 83(2)\\
     C(1) & 2980(30) & 10000 & 5000 & 1 & 86(4)\\
     C(2) & 2890(20) & 5000 & 5000 & 1 & 87(4)\\
     N(1) & 2500(30) & 10987(2) & 5000 & 1 & 86(4)\\
     N(2) & 2410(20) & 5000 & 4013(2) & 1 & 87(4)\\
     \hline
     \multicolumn{6}{c}{*U$_{eq}$ is defined as one-third of the trace of the orthogonalized U$_{ij}$ tensor.}
   \end{tabular*}
 \end{table}

 % Anisotropic displacement parameters (Å$^2$x$10^3$) for \FPB\ at 100~K with estimated standard deviations in parentheses.
 \begin{table}[h]
     \caption{Anisotropic displacement parameters (Å$^2$x$10^3$) for \FPB\ at 100~K (\textit{Immm} space group) with estimated standard deviations in parentheses.}
     \label{tbl:XRD_anisotropic}
     \center
     \vspace{-0.3cm}
   \begin{tabular*}{0.6\textwidth}{@{\extracolsep{\fill}}lcccccc}
     \hline 
     Label & U$_{11}$ & U$_{22}$ & U$_{33}$ & U$_{12}$ & U$_{13}$ & U$_{23}$\\
     \hline
     Pb(1) & 9(1) & 15(1) & 13(1) & 0 & 0 & 0(1)\\
     Br(1) & 66(2) & 8(1) & 49(2) & 0 & 0 & 0\\
     Br(2) & 10(1) & 106(3) & 116(4) & -3(2) & -11(2) & -1(1)\\
     Br(3) & 91(2) & 60(2) & 7(1) & 0 & 0 & 0\\
     Br(4) & 115(3) & 53(2) & 15(1) & 0 & 0 & 0\\
     Br(5) & 170(5) & 14(2) & 65(2) & 0 & 0 & 0\\
     \hline
     \multicolumn{7}{c}{The anisotropic displacement factor exponent}\\
     \multicolumn{7}{c}{takes the form: $-2\pi2[h^2a^{*2}U_{11} +...+ 2hka^*b^*U_{12}]$.}
   \end{tabular*}
 \end{table}

 % Bond lengths (Å) for \FPB\ at 100~K with estimated standard deviations in parentheses.
 \begin{table}[h]
     \caption{Bond lengths (Å) for \FPB\ at 100~K (\textit{Immm} space group) with estimated standard deviations in parentheses.}
     \label{tbl:XRD_bond_lengths}
     \center
     \vspace{-0.3cm}
   \begin{tabular*}{0.3\textwidth}{@{\extracolsep{\fill}}ll}
     \hline 
     Label & Distances (Å)\\
     \hline
     Pb(1)-Br(1) & 2.9787(5)\\
     Pb(1)-Br(2) & 2.96058(15)\\
     Pb(1)-Br(3) & 2.9726(5)\\
     Pb(1)-Br(4) & 2.9741(5)\\
     Pb(1)-Br(5) & 2.9692(5)\\
     C(1)-N(1) & 1.300(18)\\
     C(1)-N(2) & 1.300(17)\\
     \hline
   \end{tabular*}
 \end{table}

 % Bond angles (°) for \FPB\ at 100~K with estimated standard deviations in parentheses.
 \begin{table}[h]
     \caption{Bond angles (°) for \FPB\ at 100~K (\textit{Immm} space group) with estimated standard deviations in parentheses.}
     \label{tbl:XRD_bond_angles}
     \center
     \vspace{-0.3cm}
   \begin{tabular*}{0.3\textwidth}{@{\extracolsep{\fill}}ll}
     \hline 
     Label & Angles (\degs)\\
     \hline
     Pb(1)-Br(1)-Pb(1)' & 167.51(9)\\
     Pb(1)-Br(2)-Pb(1)' & 180\\
     Pb(1)-Br(3)-Pb(1)' & 168.41(12)\\
     Pb(1)-Br(4)-Pb(1)' & 168.76(13)\\
     Pb(1)-Br(5)-Pb(1)' & 170.75(14)\\
     \hline
   \end{tabular*}
 \end{table}

\clearpage

\section{\label{SI_sec_PO}Polarization-Orientation (PO) Raman scattering measurements}
\begin{figure*}[hb]
        \centering
        \includegraphics[scale=0.4]{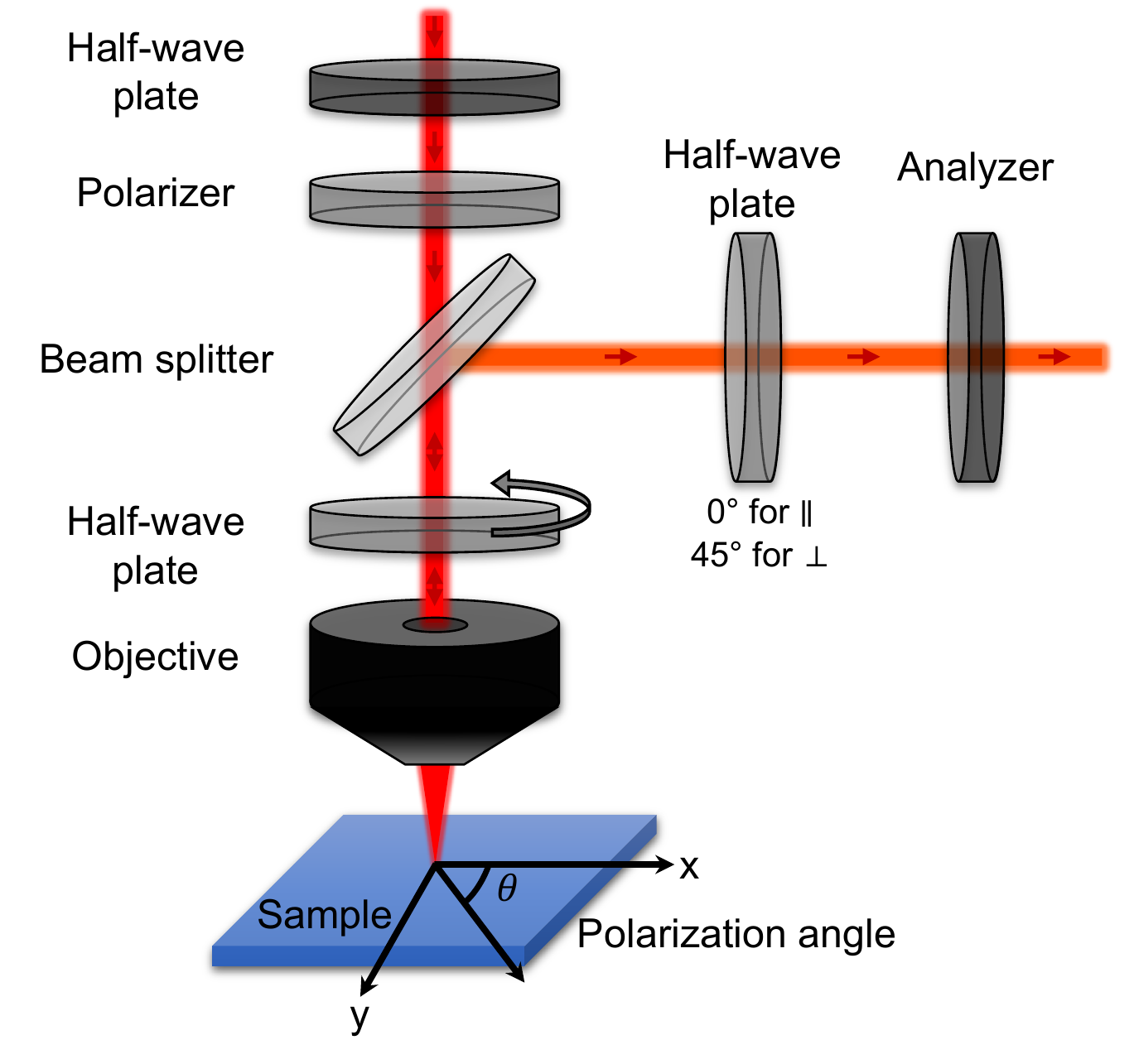}
        \caption{Scheme of the polarization-orientation Raman scattering measurements, detailing the optical components in the setup.}
        \label{SI_PO_setup}
    \end{figure*}

\begin{figure*}[hb]
            \centering
            \includegraphics[scale=0.65]{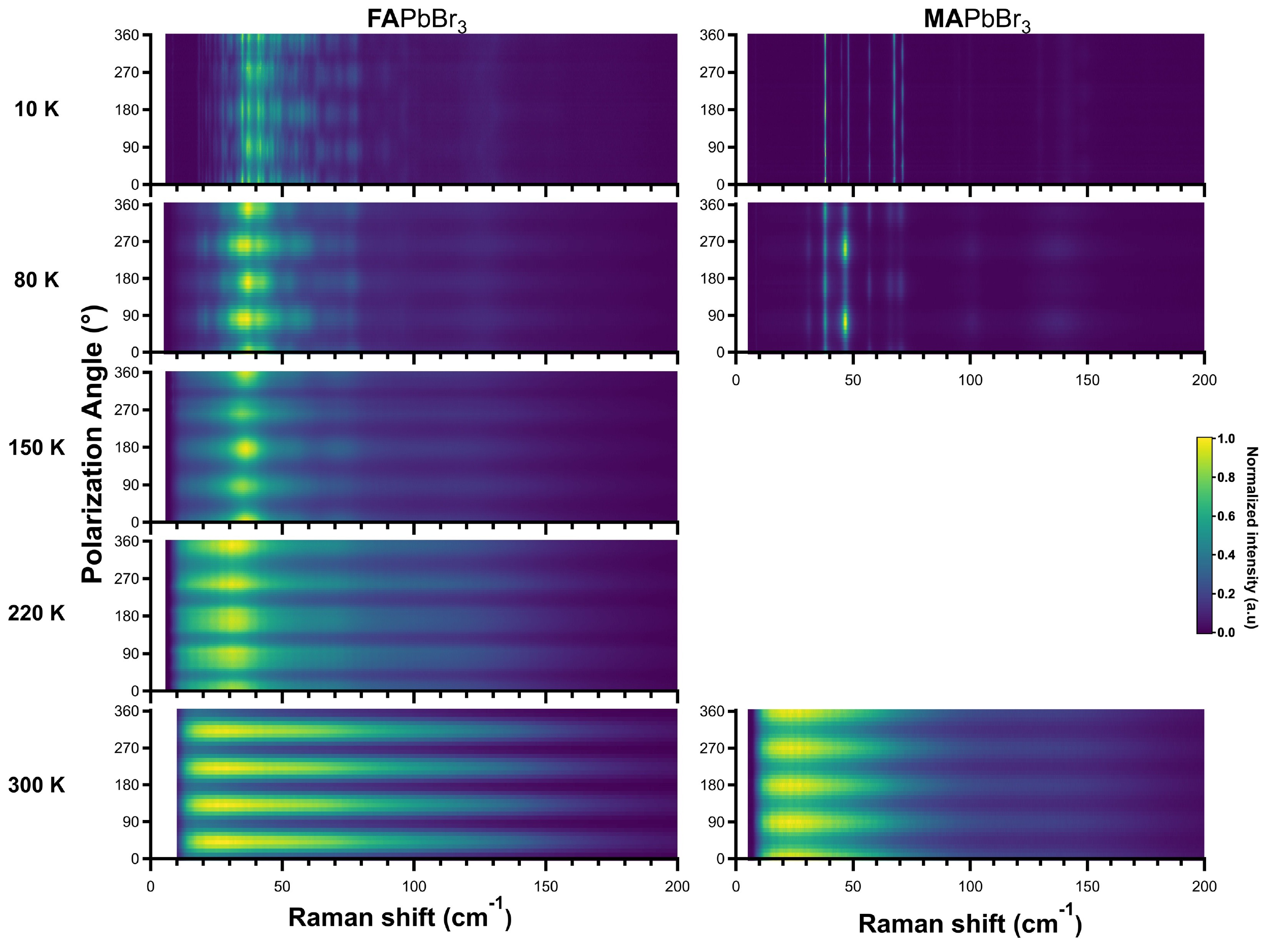}
            \caption{Temperature-dependent, false-color polarization-orientation Raman plots of \FPB\ and \MPB.
            All plots are presented for the parallel configuration.
            Intensities are separately normalized for each temperature, represented by the color scale.}
            \label{SI_PO_Raman_temperatures}
        \end{figure*}

    In polarization-orientation (PO) Raman scattering measurements, the crystal surface is excited by a linearly polarized laser (785 nm). 
    The scattered light is then filtered by an analyzer for polarization parallel and perpendicular to the incident light.
    This measurement is repeated after rotating the polarization of the incident light by 5\degs\ (half-wave plate is rotated by 2.5\degs) while the sample position is fixed (see Fig.~\ref{SI_PO_setup}).
    The resulting false-color plots (fig.~\ref{XRD_PO}(a) in the main article and fig.~\ref{SI_PO_Raman_temperatures}) show the fluctuations in scattering intensity as a function of the angle between the polarization of the incident light and an arbitrary axis on the surface of the measured crystal.
    Figure~\ref{SI_PO_Raman_temperatures} shows the temperature-dependent PO plots of \FPB\ and \MPB\ at 10~K, 80~K, and 300~K for both materials, and additionally at 150~K and 220~K for \FPB\ only.
    
\clearpage
\section{\label{SI_sec_methods}Methods}
    \subsection{Raman scattering}
        Raman scattering measurements were conducted in a home-built back-scattering system\cite{asher2020anharmonic,sharma2020lattice} using a below band-gap 1.58 eV CW pump-diode laser (Toptica Inc., USA).
        The incident beam was linearly polarized by a Glan-laser polarizer (Thorlabs, USA), directed into a microscope (Zeiss, USA), and focused on the sample through a 0.55 NA/50x objective (Zeiss, USA).
        The excitation polarization was controlled by a zero-order half-wave plate (Thorlabs, USA) and was rotated in small increments (5\degs) between measurements.
        The back-scattered beam was collected by the objective and passed through another polarizer to collect only light that was scattered either parallel or perpendicular to the incident polarization.
        Rayleigh scattering was reduced by passing the beam through a volume holographic beam-splitter and two OD$>$4 notch filters (Ondax Inc., USA).
        Finally, the beam was focused to a 1 m long spectrometer (FHR 1000, Horiba) dispersed by 1800 gr/mm grating, achieving $\approx$ 0.3~ cm$^{-1}$ spectral resolution, and detected by Si CCD (Horiba Inc., USA).
        For temperature control, all crystals were mounted into liquid Nitrogen- or liquid Helium-cooled optical cryostat (Janis Inc., USA).
        Unpolarized spectra were obtained by summing the spectra of all measured incident polarizations collected both in parallel and perpendicular configurations and normalizing them to the maximum intensity.
        The temperature-dependent Raman results presented in the main text were reversible with temperature.

    % \subsection{Single-crystal X-ray diffraction}

\subsection{First-principles calculations}
The vibrational density of states of cubic \FPB\ and \MPB\ was calculated with the finite-displacement method, using density functional theory (DFT) and the phonopy-package\cite{togo_first_2015}. 
The DFT calculations were done with the VASP package\cite{kresse1996efficient,kresse_efficiency_1996}, employing the PBE functional\cite{perdew_generalized_1996}, PAW pseudopotentials\cite{kresse_ultrasoft_1999} and a TS-scheme to correct for dispersive contributions\cite{tkatchenko_accurate_2009}. 
The energy threshold was set to $10^{-8}$~eV, and a $\Gamma$-centered k-point grid of 4x4x4 k-points was used for the 2x2x2 supercells. 
The plane-wave cutoff was set to 700~eV for \FPB\ and 800~eV for \MPB.
    We computed the cubic phase of the crystals because we found the energetic landscape of orthorhombic (\textit{i.e.}, low-temperature phase) \FPB\ to be highly sensitive to small geometrical changes of the FA molecule and associated distortions of the inorganic lattice.
    The result of this is a considerably "corrugated" potential energy surface with many essentially isoenergetic local minima, which prevents us from obtaining a numerically stable, well-defined equilibrium structure of the \FPB\ orthorhombic crystal that would be a prerequisite for subsequent phonon calculations.

%\bibliography{literature.bib}\clearpage
% \input{main.bbl}
% %\input{Supporting_Information.bbl}
% \input{main_combined.bbl}

%apsrev4-2.bst 2019-01-14 (MD) hand-edited version of apsrev4-1.bst
%Control: key (0)
%Control: author (8) initials jnrlst
%Control: editor formatted (1) identically to author
%Control: production of article title (0) allowed
%Control: page (0) single
%Control: year (1) truncated
%Control: production of eprint (0) enabled
%

%\twocolumngrid

% \section*{References}
% \bibliography{literature.bib}

\end{document}